\documentclass[a4paper]{article}
\usepackage{amsmath,amssymb}

\newcommand{\dd}{{\rm d}} \newcommand{\e}{{\rm e}} \newcommand{\im}{{\rm i}}
\newcommand{\Tr}{{\rm Tr}} \renewcommand{\Im}{{\rm Im}} \renewcommand{\Re}{{\rm Re}}
\newcommand{\bbbone}{{\mathchoice {\rm 1\mskip-4mu l} {\rm 1\mskip-4mu l}
{\rm 1\mskip-4.5mu l} {\rm 1\mskip-5mu l}}}
\renewcommand{\rho}{\varrho} \renewcommand{\theta}{\vartheta}
\renewcommand{\epsilon}{\varepsilon} \renewcommand{\kappa}{\varkappa}
\renewcommand{\phi}{\varphi}

 \newcommand{\Fscr}{{\mathcal F}}
 \newcommand{\Hscr}{{\mathcal{H}}}
 \newcommand{\Lscr}{{\mathcal{L}}}

 \newcommand{\Uscr}{{\mathcal{U}}}
 \newcommand{\Wscr}{{\mathcal{W}}}
\newcommand{\Xscr}{{\mathcal{X}}} 
\newcommand{\Zscr}{{\mathcal{Z}}}
\newcommand{\Proof}{{\smallskip\bf \noindent Proof.\ }}
\newcommand{\finedim}{{\nobreak\kern5pt\nobreak\vrule height4pt width4pt depth0pt
\vskip4pt plus2pt}}

\newtheorem{theorem}{Theorem}
\newtheorem{proposition}[theorem]{Proposition}

\begin{document}



\title{Photoemissive sources and \\
quantum stochastic calculus}

\author{A.~Barchielli \\
{\small  {\sl Dipartimento di Matematica, Politecnico di Milano, Piazza Leonardo da Vinci 32,}}
\\ {\small {\sl I-20133 Milano, Italy} and {\sl Istituto Nazionale di Fisica Nucleare,
Sezione di Milano}}{\small } \\
\\ G.~Lupieri\\  {\small  {\sl Dipartimento di Fisica, Universit\`a degli Studi di
Milano,
Via Celoria 16,}} \\ {\small {\sl I-20133 Milano, Italy}  and {\sl Istituto Nazionale di Fisica
Nucleare, Sezione di Milano}}}

\date{1991 {\sl Mathematics Subject Classification}: Primary 81S25; Secondary 81V80.}
\maketitle

\section{Introduction.}

Just at the beginning of quantum stochastic calculus (QSC), Hudson and Parthasarathy proposed a
quantum stochastic Schr{\"o}dinger equation linked to dilations of quantum dynamical semigroups
\cite{1HP84,2HP84,Partha92}. Such an equation has found applications in physics, mainly in
quantum optics, but not in its full generality \cite{Gar, Bar90, Bar93}. It has been used to
give, at least approximately, the dynamics of photoemissive sources such as an atom absorbing
and emitting light or matter in an optical cavity, which exchanges light with the surrounding
free space. But in these cases the possibility of introducing the gauge (or number) process in
the dynamical equation has not been considered. In this paper we want to show, in the case of
the simplest photoemissive source, namely a two--level atom stimulated by a laser, how the full
Hudson--Parthasarathy equation allows to describe in a consistent way not only absorption and
emission, but also the scattering of the light by the atom.

Let us recall the Hudson--Parthasarathy equation; this is just to fix our notations, while for
the proper mathematical definitions and the rules of QSC we refer to the book by Parthasarathy
\cite{Partha92}. We denote by $\Fscr := \Fscr(\Xscr)$ the Boson Fock space over the Hilbert
space $\Xscr := \Zscr \otimes L^2({\mathbb R}_+) \simeq L^2({\mathbb R}_+ ; \Zscr)$, where
$\Zscr$ is another separable complex Hilbert space. Let $\{e_i,\ i\geq 1\}$ be a c.o.n.s. in
$\Zscr$ and let us denote by $A_i(t)$, $A^\dagger_i(t)$, $\Lambda_{ij}(t)$ the annihilation,
creation and gauge processes associated with such a c.o.n.s. We denote by $E(h)$, $h\in \Xscr$,
the exponential vectors in $\Fscr$ with normalization $\| E(h)\|^2 = \exp\{ \|h\|^2\}$; $E(0)$
is the Fock vacuum. We shall also use  the Boson Fock spaces $\Fscr_t := \Fscr\big( L^2
([0,t];\Zscr)\big)$ and $\Fscr^t := \Fscr\big( L^2 ((t,\infty);\Zscr)\big)$, for which we have
$\Fscr= \Fscr_t\otimes \Fscr^t$, and the Weyl operators
\begin{equation}
\Wscr_t(f):= \exp \left\{ \sum_j\int_0^t \left[f_j(s)\, \dd A_j^\dagger(s) - \overline{ f_j(s)}
 \, \dd A_j(s) \right] \right\}, \qquad f\in L^2_{{}_{\rm loc}}({\mathbb R}_+;\Zscr)\,.
\end{equation}

Let $\Hscr$ be a separable complex Hilbert space (the system space) and let $H_0$, $\{R_i^0$,
$i\geq 1\}$, $\{S_{ij},\ i,j\geq 1\}$, be bounded operators in $\Hscr$ such that $H_0^*=H_0$,
$\sum_i R_i^{0*} R_i^0$ is strongly convergent to a bounded operator, and $\sum_{i,j} S_{ij}
\otimes |e_i \rangle \langle e_j|=: S\in \Uscr(\Hscr\otimes \Zscr)$ (unitary operators in
$\Hscr \otimes \Zscr$); we set also
 $K_0:=H_0 - \frac{\im}{2} \sum_j R_j^{0*}R_j^0$.
Then (\cite{Partha92} Theor.~27.8 p.~228) there exists a unique unitary operator--valued
adapted process $U(t)$ satisfying $U(0) =\bbbone$ and
\begin{equation}\label{1.2}
\dd U(t) = \bigg\{ \sum_j R_j^0 \,\dd A_j^\dagger(t) + \sum_{i,j} \left(S_{ij}-
\delta_{ij}\right) \dd \Lambda_{ij}(t)
- \sum_{i,j} R_i^{0*} S_{ij }\,\dd A_j(t) -\im K_0\,\dd t \bigg\} \, U(t)\,.
\end{equation}

Now, $\Fscr$ is interpreted as the Hilbert space of the electromagnetic field; $A^\dagger_j(t)$
creates a photon of type $j$ in the time interval $[0,t]$, $A_j(t)$ annihilates it,
$\Lambda_{jj}(t)$ is the selfadjoint operator representing the number of photons of type $j$ up
to time $t$ and $N(t):= \sum_j \Lambda_{jj}(t)$ is the observable ``total number of photons up
to time $t$". We shall see in Section 3 how to choose the one--particle space $\Zscr$
[eq.~(\ref{3.22})].

In order to describe a two--level atom, we take $\Hscr= {\mathbb C}^2$; then, to fix the model,
we have to determine the atomic operators $H_0$, $R_i^0$, $S_{ij}$ on the basis of physical
considerations. Let us note that, if $R_i^0=0$, the flux of incoming photons turns out to be
equal to that of outgoing ones, but in general, for an arbitrary choice of the system
operators, the flux conservation does not hold.  Physically, flux conservation is to be
expected when the possible processes are absorption/emission and elastic scattering; note that
in the presence of absorption/emission this conservation cannot be instantaneous, but only in
the average over long times. In the next section we shall require a weak form of photon flux
conservation, namely only in the mean and for large times [eqs.~(\ref{x.7}), (\ref{2.2})]. This
suffices to determine the structure of the atomic operators [eqs.~(\ref{2.3}), (\ref{2.4})] and
to eliminate anelastic scattering; an interesting balance equation [eq.~(\ref{2.23})] is
obtained as a byproduct. We end Section 2 by studying the large--time behaviour of the reduced
atomic density matrix [eqs.~(\ref{2.21})--(\ref{3.2})].

In Section 3, in order to give an example of physical consequences of the model obtained in
Section 2, we study the differential and total cross sections for the scattering of laser light
by the atom, as a function of the frequency of the stimulating laser [eqs.~(\ref{3.14}),
(\ref{3.15})]. The resulting line--shape is very interesting. Not only a Lorentzian shape is
permitted, but the full variety of Fano profiles (\cite{CT92} pp.~61--63) can be obtained
[eq.~(\ref{3.16})]. Moreover, the dependence of the line shape on the intensity of the
stimulating laser is computed; in particular, the resonance position turns out to be intensity
dependent [eq.~(\ref{3.26})], a phenomenon known as ``lamp shift" \cite{Kast}.

\section{Master equation and flux conservation.}
First of all we want a model for an atom stimulated by a laser; this means to choose as initial
state $\Psi\in \Hscr\otimes \Fscr$ a generic state for the atom and a coherent vector for the
field \cite{Bar90}, i.e.
\begin{equation}\label{x.1}
  \Psi = \xi \otimes \frac{E(f)}{\|E(f)\|}\,, \qquad \xi \in \Hscr\,, \quad \|\xi\|=1\,,
  \quad f\in L^2({\mathbb R}_+;\Zscr)\,.
\end{equation}
Moreover, we shall consider only adapted observables $X_t \in \Lscr(\Hscr \otimes \Fscr_t)$ for
which we have
\begin{equation}\label{x.2}
  \langle U(t) \Psi | X_t\, U(t) \Psi \rangle =
  \langle \widehat U(t) \xi \otimes E(0) | \Wscr_t(f)^*X_t \Wscr_t(f)\, \widehat U(t)
  \xi\otimes E(0) \rangle\,,
\end{equation}
\begin{equation}\label{x.3}
  \widehat U(t) := \Wscr_t(f)^*U(t) \Wscr_t(f)\,.
\end{equation}
This implies that in all physical expressions we can take $f\in L_{{}_{\rm loc}}^2({\mathbb
R}_+;\Zscr)$ and indeed, in order to describe monochromatic coherent light, we choose
\begin{equation}\label{x.4}
  f(t) = \e^{-\im \omega t}\, \lambda\,, \qquad \lambda\in \Zscr\,, \quad \omega\in {\mathbb R}
  \,.
\end{equation}

 Let us recall that the atomic reduced statistical operator $\rho_\lambda(t)$ is
defined by the partial trace
\begin{equation}\label{x.5}
  \rho_\lambda(t) := \Tr_{{}_\Fscr} \left\{ U(t) |\Psi \rangle \langle \Psi| U(t)^* \right\}
  \equiv \Tr_{{}_\Fscr} \left\{ \widehat U(t) |\xi\otimes E(0) \rangle \langle \xi\otimes
  E(0)| \widehat U(t)^*
  \right\}\,.
\end{equation}
Moreover, the quantity
\begin{equation}\label{x.6}
  \langle N(t) \rangle := \langle U(t) \Psi | N(t)\, U(t) \Psi \rangle
  \equiv \langle\widehat U(t) \xi\otimes E(0)|\Wscr_t(f)^* N(t) \Wscr_t(f) \widehat U(t) \xi\otimes
  E(0)\rangle
\end{equation}
represents the mean number of photons up to time $t$, after the interaction with the atom,
while $\langle \Psi | N(t)\, \Psi\rangle$ is the same quantity before such an interaction
\cite{Bar90}. By the theory of quantum continuous measurements \cite{BarL85, Bar90} the
probability law of the counting stochastic process associated with the observables $N(t)$,
$t\geq 0$, could be obtained; however, in this paper we shall need only mean values such as
(\ref{x.6}) and not the full theory of continuous measurements.

In order to formulate physical requirements, let us start by considering the case when no
photon is injected into the system, i.e. $\lambda=0$. In these conditions it is natural to ask
that the atom can emit or one or zero photons depending on the atomic initial state; moreover,
we ask the final state to be independent from the initial one; this is done in the next
proposition.

\begin{proposition}
We assume that
\begin{equation}\label{x.7}
  \lambda=0 \quad \Longrightarrow \quad\forall \xi\,, \ \forall t\,, \quad \langle N(t)\rangle
  \leq 1\,,
  \quad \rho_0(t) \stackrel{t \to +\infty}{ \longrightarrow } \rho_{{}_{\rm eq}}^0\,;
\end{equation}
moreover, we take as canonical basis $\{|+\rangle,\,|-\rangle\}$ in $\Hscr$ the basis which
diagonalises $\rho_{{}_{\rm eq}}^0$, so that we can write $\rho_{{}_{\rm eq}}^0= pP_+
+(1-p)P_-$ for some $p$ in $[0,1]$. Then, apart from an exchage of roles between the two
states $|+\rangle$, $|-\rangle$, we obtain $\rho_{{}_{\rm eq}}^0= P_-$ and
\begin{equation}\label{2.3}
  H_0= \frac 1 2 \, \omega_0 \sigma_z\,, \quad \omega_0 \in {\mathbb R}\,, \qquad
  R_j^0 = \langle e_j|\alpha \rangle\, \sigma_-\,, \quad \alpha \in \Zscr \,,
  \ \alpha\neq 0\,.
\end{equation}
\end{proposition}

Let us recall that $\sigma_z$, $\sigma_\pm$ are the Pauli matrices, which are given by
\begin{equation}\label{2.1}
\sigma_z := \left(\begin{array}{cc} 1& 0 \\ 0 &-1 \end{array}\right),
\qquad \sigma_+ := \left(\begin{array}{cc} 0& 1 \\ 0 &0 \end{array}\right),
\qquad \sigma_- := \left(\begin{array}{cc} 0& 0 \\ 1 &0 \end{array}\right);
\end{equation}
moreover, the two orthogonal projections $P_\pm$  can be written as
$
P_+ = \frac 1 2 (\bbbone + \sigma_z) = \sigma_+ \sigma_-$, $P_- = \frac 1 2 (\bbbone -
\sigma_z) = \sigma_- \sigma_+$.

\Proof
By using the rules of QSC one obtains that $\rho_0(t)$ satisfies the master equation
\begin{equation}\label{x.8}
\frac{\dd\rho_0(t)}{ \dd t} = \Lscr_0[\rho_0(t)]\,,\qquad
  \Lscr_0[\rho] := -\im [H_0\,, \rho] +\frac 1 2 \sum_j \left(
  \left[ R^0_j \rho\,,\, R^{0 *}_j \right] + \left[ R^0_j\,,\, \rho
  R^{0 *}_j \right]\right),
\end{equation}
and that
\begin{equation}\label{x.9}
  \langle N(t) \rangle = \int_0^t \Tr_{{}_\Hscr} \Bigl\{ \sum_j R_j^{0*} R_j^0 \rho_0(s)
  \Bigr\} \dd s\,.
\end{equation}
From eqs.~(\ref{x.9}) and (\ref{x.7}), we obtain the necessary condition $
\sum_j\Tr_{{}_\Hscr}
\left\{ R_j^{0*} R_j^0 \rho_{{}_{\rm eq}}^0  \right\}=0$. By the cyclic property of the trace
and the positivity of $\rho_{{}_{\rm eq}}^0$ and $R_j^0 \rho_{{}_{\rm eq}}^0 R_j^{0*}$, we get
that this condition is equivalent to $  R_j^0\, \rho_{{}_{\rm eq}}^0 = 0 $, $\forall j$.

Now, let us set $R_j^0 = x_j\bbbone +y_j \sigma_z + z_j \sigma_+ + \alpha_j
\sigma_-$ (every operator on ${\mathbb C}^2$ can be written in this way).
Then,  $R_j^0\rho_{{}_{\rm eq}}^0=0$ gives $p(x_j+y_j)=0$, $(1-p)(x_j-y_j)=0$, $(1-p)z_j=0$,
$p\alpha_j=0$. For $p\in (0,1)$ this system gives $R_j^0=0$, which is not acceptable because
there is not a unique equilibrium state. For $p=0$ we get $x_j=y_j$ and $z_j=0$; we need also
$\sum_j|\alpha_j|^2\neq 0$ to have decay to an equilibrium state. We do not consider the case
$p=1$, because it is analogous to the previous one, apart from the exchange of $|+\rangle$ and
$|-\rangle$. Therefore we have $R_j^0 = \alpha_j \sigma_- +
\beta_j P_+$. From eqs.~(\ref{x.7}) and (\ref{x.8}) we have also $\Lscr_0[\rho_{{}_{\rm
eq}}^0]=0$, which reduces to $[H_0,\rho_{{}_{\rm eq}}^0]=0$. Because $H_0$ is selfadjoint and
defined up to a constant, we can write $H_0= \frac 12\, \omega_0 \sigma_z$, $\omega_0\in
{\mathbb R}$. Finally, by solving the master equation and computing $\langle N(t)\rangle$ from
eq.~(\ref{x.9}), we get that $\langle N(t)\rangle\leq 1$ implies $\beta_j=0$.
\finedim

Now, let us consider the case $\lambda\neq 0$. If the possible physical processes are
absorption/emission of single photons and elastic scattering, we expect that for large times
the mean number of injected photons $\langle \Psi| N(t)\, \Psi\rangle= \|\lambda\| t$ be equal
to the mean number of outgoing photons, so we require the flux conservation in the mean, as
done in the next proposition.

\begin{proposition}
We assume eq.~(\ref{2.3}) and
\begin{equation}\label{2.2}
  \lim_{t\to +\infty} \frac {\langle N(t) \rangle}{\langle \Psi | N(t)\, \Psi\rangle} =1\,,
  \qquad \forall \lambda\in \Zscr\,, \quad \forall \omega\,.
\end{equation}
Then, we have
\begin{equation}\label{2.4}
  S =  P_+\otimes S^+ + P_-\otimes S^-\,, \qquad   S^\pm  \in \Uscr (\Zscr)\,.
\end{equation}
\end{proposition}

\Proof This proof relies on some cumbersome computations; we only try to sketch it.
First of all by QSC we obtain
\begin{equation}
 \dd \widehat U(t) = \bigg\{ \sum_j R_j^\lambda(t) \,\dd A_j^\dagger(t) + \sum_{i,j} \left(
 S_{ij}-\delta_{ij}\right) \dd \Lambda_{ij}(t)
  - \sum_{i,j} R_i^{\lambda}(t)^* S_{ij }\,\dd A_j(t)-\im K_\lambda(t)\,\dd t \bigg\} \,
  \widehat U(t)\,,
\label{p.5}\end{equation}
\begin{equation}
  R_j^\lambda(t) :=\langle e_j| \alpha + F^-f(t) \rangle \sigma_- +
  \langle e_j|  F^+f(t)  \rangle \sigma_+ +\langle e_j| (S^+- \bbbone)f(t) \rangle P_+
  + \langle e_j| (S^-- \bbbone)f(t) \rangle P_-\,,
\label{p.6}\end{equation}
\begin{eqnarray}
  K_\lambda(t) :=& &\!\!\!\!\!\!\! H_0 - \im \|\lambda\|^2 + \im \left( \langle
   \lambda  | S^+ \lambda \rangle -\langle\alpha | F^- f(t) \rangle - \frac 1 2 \|\alpha\|^2
   \right) P_+ + \im \langle \lambda | S^-  \lambda \rangle \,P_- +{}\nonumber\\
  +& & \!\!\!\!\!\!\! \im  \big(  \langle \lambda |  F^+\lambda \rangle +\langle F^- \lambda
  |\lambda\rangle - \langle \alpha | S^- f(t) \rangle\big)\sigma_++ \im \langle f(t) |\alpha
  \rangle\sigma_-\,.\label{p.7}
  \end{eqnarray}
We have used the notation
\begin{equation}\label{p.4}
  S=:P_+\otimes S^+ + P_-\otimes S^- + \sigma_+ \otimes F^+ +\sigma_- \otimes F^-\,,
  \qquad S^\pm,\, F^\pm \in \Lscr(\Zscr)\,;
\end{equation}
the unitarity of $S$ implies some simple relations among $S^\pm$, $F^\pm$.

Then, we introduce the quantity
\begin{equation}\label{p.1}
  Y(t) := \langle N(t) \rangle + \frac 1 2 \, \Tr_{{}_\Hscr} \left\{\sigma_z \rho_\lambda (t)
  \right\} - \frac 12 \, \langle \xi| \sigma_z \xi \rangle - \int_0^t \|f(s)\|^2 \,\dd s\,,
\end{equation}
whose derivative can be computed by QSC from eqs.~(\ref{x.5}), (\ref{x.6}),
(\ref{p.5})--(\ref{p.7}). By using the unitarity of $S$ and the fact that $Y(0)=0$, we can
write
\begin{equation}\label{p.2}
  Y(t) = \int_0^t  \left\{ -\rho_{\,\lambda}^{++} (s) \|F^- \lambda\|^2 + 2\Re\,
  \rho_{\,\lambda}^{+-}(s)
  \langle F^+ \lambda |S^+ \lambda \rangle + \rho_{\,\lambda}^{--}(s)\|F^+ \lambda\|^2
   \right\}\dd s \,,
\end{equation}
where $\rho_{\,\lambda}^{++} (t)\,,\,\ldots$ are the matrix elements of $\rho_\lambda(t)$.

 Again by applying the rules of QSC to eqs.~(\ref{x.5}) and (\ref{p.5}), we obtain the
master equation
\begin{equation}\label{2.22}
  \frac{\dd \rho_\lambda (t)}{\dd t} = -\im K_\lambda(t) \rho_\lambda(t) +\im \rho_\lambda(t)
  K_\lambda(t)^* + \sum_j R_j^\lambda(t) \rho_\lambda(t) R_j^\lambda(t)^*\,;
\end{equation}
now, the Liouvillian is time dependent. In general an equilibrium state does not exist, but,
for large $t$, $\rho_\lambda (t)$ can contain only constant and oscillating terms; therefore,
the Cesaro limit $\lim_{t \to \infty} \frac 1 t
\int_0^t
\rho_\lambda (s)\,\dd s=:
\widehat
\rho$ exists. By dividing eq.~(\ref{p.2}) by $\|\lambda\|^2 t$ and taking the limit, we obtain
\begin{equation}
  \lim_{t\to +\infty} \frac {\langle N(t) \rangle}{\langle \Psi | N(t)\, \Psi\rangle} -1
  = \big\|F^+ \widehat \lambda \big\|^2 + 2 \, \Re\,  \widehat \rho^{+-}
   \big\langle F^+ \widehat \lambda \big| S^+ \widehat \lambda \big\rangle
  -
   \widehat \rho^{++} \Bigl(  \big\|F^+ \widehat \lambda
  \big\|^2 + \big\|F^- \widehat \lambda \big\|^2\Bigr),\label{p.3}
\end{equation}
where $\widehat \lambda := \lambda/\|\lambda\|$. We want the r.h.s. of eq.~(\ref{p.3}) to be
zero for every $\lambda$, but we know that $\widehat \rho\to P_-$ for $\|\lambda\|\to 0$ and,
in particular, we have $\widehat \rho^{++} \to 0$ and $\widehat \rho^{+-} \to 0$. Therefore,
the flux conservation implies $\|F^+\widehat\lambda\|=0$ for every $\widehat\lambda$, i.e.
$F^+=0$.

If $\Zscr$ were finite dimensional, the unitarity of $S$ and $F^+=0$ would imply directly
$F^-=0$. In the infinite dimensional case, one can only conclude that $F^{-*}F^-=:Q$ is an
orthogonal projection. Some more considerations are needed to have $F^-=0$. The line of the
proof is the following. For every $\lambda$ with a non vanishing component in the range of $Q$
the flux conservation would imply $\widehat\rho^{++}=0$ and so $\widehat\rho=P_-$. Moreover,
$P_-$ would be an invariant state for the master equation (\ref{2.22}); but to impose this
gives $F^-=0$, contrarily to our hypothesis. So eq.~(\ref{2.4}) is proved.
\finedim

From now on we assume eqs.~(\ref{x.1}), (\ref{x.4}), (\ref{2.3}), (\ref{2.4}) to hold and,
always for physical reasons, we take
\begin{equation}\label{2.20}
  \omega_0 >0\,, \qquad \omega>0\,.
\end{equation}
Let us stress that as a byproduct of the asymptotic flux conservation (\ref{2.2}) we have
obtained the balance equation
\begin{equation}\label{2.23}
  \langle N(t)\rangle +\frac 1 2 \Tr_{{}_\Hscr} \left\{ \sigma_z \bigl( \rho_\lambda(t) -
  \rho_\lambda(0) \bigr) \right\} = \|\lambda\|^2 t\,,
\end{equation}
which means that the mean number of outgoing photons plus the mean number of absorbed photons
$\big($the second term reduces to $\rho_{\,\lambda}^{++}(t)-\rho_{\,\lambda}^{++}(0)\big)$ is
equal to the mean number of ingoing photons.

Let us end this section by studying the asymptotic behaviour of $\rho_\lambda(t)$. By setting
\begin{equation}\label{2.5}
  \widetilde \rho_\lambda(t) := \exp \left\{ \im \omega \sigma_z t/2\right\} \rho_\lambda(t)\,
  \exp \left\{ -\im \omega \sigma_z t/2\right\},
\end{equation}
we obtain, by QSC, the master equation
\begin{equation}\label{2.6}
  \frac{\dd \ }{\dd t}\, \widetilde\rho_\lambda(t) = \Lscr_\lambda \left[\widetilde\rho_\lambda
  (t)\right],
\end{equation}
with
\begin{equation}\label{2.17}
  \Lscr_\lambda[\rho] := -\im [H_\lambda\,, \rho] +\frac 1 2 \sum_j \left(
  \left[\widetilde R^\lambda_j \rho\,,\,\widetilde R^{\lambda *}_j \right] + \left[\widetilde
   R^\lambda_j\,,\, \rho\widetilde  R^{\lambda *}_j \right]\right),
\end{equation}
\begin{equation}\label{2.18}
  \widetilde R_j^\lambda := \langle e_j | \alpha \rangle \sigma_- + \langle e_j |(S^+ -\bbbone)
  \lambda \rangle P_+  +\langle e_j |(S^- -\bbbone)  \lambda \rangle P_-\,,
\end{equation}
\begin{eqnarray}
  H_\lambda :=& &\!\!\!\!\!\!\! \frac 12\, (\omega_0 - \omega)\sigma_z -\Im \,\langle \lambda
  |  S^+ \lambda
  \rangle P_+-\Im \,\langle \lambda | S^- \lambda \rangle P_- +{} \nonumber \\
  {}+& &\!\!\!\!\!\!\!\frac \im 2 \langle (
  S^- + \bbbone ) \lambda |\alpha \rangle \sigma_- - \frac \im 2 \langle \alpha |(
  S^- + \bbbone ) \lambda \rangle \sigma_+\,.
\label{2.19}\end{eqnarray}

The general master equation for a two--level system is studied in \cite{Lendi}; in the
following we shall use similar techniques, apart from a different parametrization of the
statistical operator which turns out to be more convenient in our case. By setting
\begin{equation}\label{2.8}
  \widetilde\rho_\lambda(t) =: \left( \begin{array}{cc} u(t) &  v(t) \\  & \\
  \overline{v(t)} &  1-u(t) \end{array} \right) \qquad \qquad\left\{ \begin{array}{l} 0\leq u(t)
  \leq 1  \\ \\ u(t) \geq u^2(t) +  | v(t)|^2 \end{array}\right.
\end{equation}
where the conditions on the right express the fact that $\widetilde\rho_\lambda (t)$ is a
statistical operator, we obtain from the master equation
\begin{equation}\label{2.13}
  \frac{\dd \ }{\dd t}\, {\mathbf u}(t) =  {\mathbf G\, u }(t) - {\mathbf w}\,,
\end{equation}
where
\begin{equation}\label{2.14}
  {\mathbf u}(t) := \left( u(t), v(t), \overline{v(t)} \right)^{\rm T}, \qquad
   {\mathbf w} := \left( 0, \left\langle \alpha | S^- \lambda \right\rangle
   ,  \left\langle  S^- \lambda | \alpha\right\rangle\right)^{\rm T}\,,
\end{equation}
\begin{equation}\label{2.9}
 {\mathbf G}:= \left( \begin{array}{ccc} -\|\alpha\|^2 & - \left \langle S^- \lambda
 | \alpha \right\rangle & -\left \langle \alpha |
 S^- \lambda \right \rangle \\ & & \\
 \left \langle \alpha | \left( S^+ +S^-\right) \lambda \right \rangle &
 -b & 0 \\ & & \\
 \left \langle \left( S^+ +S^-\right) \lambda | \alpha \right \rangle & 0
 & - \overline b \end{array} \right),
\end{equation}
\begin{equation}\label{2.12}
  b:= \frac 1 2 \, \kappa^2 - \im \left( \Delta \omega - \Im \left \langle S^+ \lambda
  | P_\alpha  S^- \lambda \right\rangle \right),
\end{equation}
\begin{equation}\label{2.11}
   \kappa^2 := \mu^2 + \left| \left \langle \widetilde \alpha
  | \Delta S \lambda \right \rangle  \right | ^2 \,, \qquad
  \mu^2 := \|\alpha\|^2 + \|P_\bot \Delta S\lambda\|^2\,,\qquad
  \Delta S := S^+ -S^-\,,
\end{equation}
\begin{equation}\label{2.10}
  \Delta \omega := \omega - \left( \omega_0 + \Im \left\langle S^+ \lambda | P_\bot S^-
  \lambda\right\rangle \right),
\end{equation}
\begin{equation}\label{2.7}
  \widetilde \alpha := \frac  \alpha{\|\alpha\|}\,, \qquad P_\alpha := \left| \widetilde \alpha
  \right\rangle \left \langle \widetilde \alpha \right|, \qquad P_\bot := \bbbone - P_\alpha\,.
\end{equation}
Moreover, we have
\begin{equation}\label{2.15}
  {\rm det}\, {\mathbf G} = - \|\alpha\|^2 \left[ \left( \Delta \omega \right)^2 +
  \Gamma^2/4 \right],
\end{equation}
with
\begin{eqnarray}
  \Gamma^2 :=& & \!\!\!\!\!\!\!\kappa^4 +4 \kappa^2 \,\Re \left\langle S^- \lambda | P_\alpha \left( S^+ +
  S^- \right) \lambda \right \rangle - 4 \left( \Im \left\langle  S^+ \lambda | P_\alpha
  S^- \lambda \right \rangle \right)^2 \equiv \label{2.16}\\
  \equiv & & \!\!\!\!\!\!\!\left( \mu^2 + 2 \left| \left\langle \widetilde \alpha | S^- \lambda \right
  \rangle \right |^2 - 2\,\Re \left \langle S^+ \lambda | P_\alpha S^- \lambda \right \rangle
  \right)^2
  +\left| \left \langle \widetilde \alpha \big| \left( S^+ + S^- \right) \lambda
  \right\rangle \right|^2 \left( 2 \mu^2 + \left| \left \langle \widetilde \alpha \big|  \Delta
  S \lambda \right \rangle \right|^2 \right).\nonumber
\end{eqnarray}
Let us note that $\|\alpha\|>0$ implies ${\rm det}\, {\mathbf G} < 0$ and $\Gamma^2 >0$.
Finally, the equilibrium state is given by
\begin{equation}\label{2.21}
  \lim_{t\to +\infty} \widetilde \rho_\lambda(t) =: \rho^\lambda_{\rm eq} =
  \left( \begin{array}{cc} u(\infty) & v(\infty) \\
  {}&{}\\
  \overline{v(\infty)} & 1-u(\infty) \end{array} \right),
  \end{equation}
where $u(\infty)$ and $v(\infty)$ are computed by equating to zero the time derivative in
eq.~(\ref{2.13}); then, we have ${\mathbf u}(\infty)= {\mathbf G}^{-1} {\mathbf w}$, which
gives
\begin{equation}\label{3.1}
  u(\infty) = \frac {\kappa^2 \|P_\alpha S^- \lambda\|^2}{ (\Delta \omega)^2 + \Gamma^2/4}\,,
\end{equation}
\begin{equation}\label{3.2}
  v(\infty) = {}- \frac {\langle \alpha| S^- \lambda \rangle}{(\Delta \omega)^2 + \Gamma^2/4}
  \left( \frac{\kappa^2} 2 + \im \Delta \omega +\im \,\Im\, \langle S^+ \lambda | P_\alpha
  S^- \lambda \rangle \right).
\end{equation}

Quantities like $\omega_0$, $\alpha$, $S^\pm$ are phenomenological parameters, or, better, they
have to be computed from some more fundamental theory, such as some approximation to quantum
electrodynamics. The whole model is meaningful only for $\omega$ not too ``far" from $\omega_0$
and $\omega_0$ must include the Lamb shifts. In the final results one can admit a slight
$\omega$-dependence in the elastic scattering matrices $S^\pm$.

\section{Cross section.}
The approximations which allows to describe the electromagnetic field as a Boson field in our
Fock space $\Fscr$ are known as \emph{quasimonochromatic paraxial approximation} (see
\cite{Bar90} and references therein). In particular in this approximation the fields behave as
monodimensional waves, so that a change of position is equivalent to a change of time and
viceversa. Moreover, we do not take into account the polarization degrees of freedom.
Therefore, the space $\Zscr$ has to contain only the degrees of freedom linked to the direction
of propagation, so that we can take
\begin{equation}\label{3.22}
  \Zscr = L^2 (\Upsilon, \dd _2 \sigma)\,, \quad \Upsilon= \{0\leq \theta \leq \pi, \ 0
  \leq \phi < 2\pi\} \,, \quad \dd _2\sigma =\sin \theta\, \dd \theta\,\dd \phi\,.
\end{equation}

In this section we want to compute the cross section, when the stimulating laser is well
collimated and the intensity of the light is detected in directions different from the
direction of propagation of the laser beam. So, to have a laser beam propagating along the
direction $\theta=0$, we take
\begin{equation}\label{3.3}
  \lambda =\eta \,\e^{\im \delta} \widetilde \lambda\,, \qquad \eta>0\,, \qquad
  \delta\in [0,2\pi)  \,,
\qquad
  \widetilde \lambda (\theta,\phi) = \frac{ 1_{[0,\Delta \theta]}(\theta)}
  {\Delta \theta \sqrt{2\pi (1-\cos \Delta \theta)}} \,;
\end{equation}
in all the physical quantities the limit $\Delta \theta \downarrow 0$ will be taken. Note that
$\|\lambda\|=\eta/ {\Delta\theta}$, because we need a not vanishing atom--field interaction in
the limit.

Let us consider a spherically symmetric atom. Then we have
\begin{equation}\label{3.6}
  \widetilde \alpha = 1/\sqrt{4\pi}\,, \qquad
  S^\pm \alpha = \e^{\im s_\pm} \alpha\,, \qquad S^{\pm *}\alpha = \e^{-\im s_\pm} \alpha
  \,,
\end{equation}
where $s_+$ and $s_-$ are the $s$-wave phase shifts for the elastic scattering in the up and
down atomic states respectively. Moreover, we set
\begin{equation}\label{3.5}
  g_\pm(\theta) := \lim_{\Delta\theta \downarrow 0}\left( P_\bot \left( S^\pm -
  \bbbone \right) \widetilde \lambda \right)
  (\theta,\phi)\,, \qquad \Delta g := g_+ - g_-\,,\qquad \Delta s := s_+ - s_-\,;
\end{equation}
the fact that $g_+$ and $g_-$ do not depend on $\phi$ is again due to spherical symmetry. Now,
in the limit $\Delta\theta \downarrow 0$, we can compute the various quantities introduced in
the previous section; by using
$
  \big\langle \widetilde \alpha \,\big |\, \widetilde \lambda \big \rangle = 1/2$ and
$
  \big\langle \widetilde \alpha \,\big |\, \left( S^\pm - \bbbone \right)\widetilde \lambda
  \big \rangle = \left( \e^{\im s_\pm} - 1\right)/2
$, we obtain
\begin{equation}\label{3.9}
  \kappa^2 = \mu^2 + \frac 12\, \eta^2 [1-\cos (\Delta s)]\,, \qquad \mu^2 =
  \|\alpha\|^2 +\eta^2 \|\Delta g\|^2\,,
\end{equation}
\begin{equation}\label{3.10}
  \Gamma^2 = \mu^4 + 2\mu^2 \eta^2 + \frac 12 \, \eta^4 [1-\cos(\Delta s)]\,,
\end{equation}
\begin{equation}\label{3.11}
  \Delta \omega = \omega - [ \omega_0 + \eta^2 \, \Im \, \langle g_+| g_- \rangle +
  \sqrt\pi \, \eta^2\, \Im \, \Delta g(0)]\,,
\end{equation}
\begin{equation}\label{3.12}
  u(\infty) = \frac{ \kappa^2 \eta^2 /4}{ (\Delta \omega)^2 + \Gamma ^2/4}\,,
\end{equation}
\begin{equation}\label{3.13}
  v(\infty) ={} - \frac 12 \, \e^{\im(\delta+s_-)}\, \frac{\|\alpha\| \eta}
  { (\Delta\omega)^2 +\Gamma^2/4} \left\{ \frac{\kappa^2}{2} +\im \left[ \Delta \omega
  - \frac{\eta^2} 4 \, \sin (\Delta s) \right] \right\}\,.
\end{equation}

By direct detection, it is possible to measure the intensity of the light (or to count the
photons) propagating in a small solid angle $\Delta \Upsilon$ around some direction
$(\theta^\prime,
\phi^\prime)$, different from the direction $\theta=0$. The observable ``number of photons in
$\Delta\Upsilon$ up to time $t$" is represented by
\begin{equation}\label{3.4}
  X_t := \sum_{i,j} \langle e_i| 1_{\delta\Upsilon} \,e_j \rangle\, \Lambda_{ij}(t)\,,
\qquad
   \langle e_i| 1_{\delta\Upsilon} \,e_j \rangle \simeq \overline{ e_i(\theta^\prime,
   \phi^\prime)} \, e_j(\theta^\prime,\phi^\prime)\, \sin \theta^\prime\, \dd \theta^\prime
   \dd \phi^\prime\,.
\end{equation}
Because of $1_{\Delta\Upsilon}\, \lambda=0$, which expresses the fact that the direction of
detection is different from the beam direction, we have $\Wscr_t(f)^* X_t \Wscr_t(f)=X_t$. By
introducing the operator ``number of photons up to time $t$ per unit of solid angle"
\begin{equation}\label{3.7}
  n(\theta,\phi;t) := \sum_{i,j}\overline{ e_i(\theta,  \phi)} \, \Lambda_{ij}(t)
  \,e_j(\theta,\phi)\,,
\end{equation}
by the previous remark and eq.~(\ref{x.2}), its mean value for $\theta\neq 0$ is given by
\begin{equation}\label{3.8}
  \langle n(\theta,\phi;t)\rangle = \big\langle  \widehat U_t \xi\otimes E(0) \,\big| \,
  n(\theta,\phi;t) \widehat U_t \xi\otimes E(0) \big \rangle\,.
\end{equation}
Then, a natural definition of differential cross section is
\begin{equation}\label{3.19}
  \sigma(\theta,\phi) = A_0 \lim_{t\to +\infty} \frac {\langle n(\theta,\phi;t) \rangle}
  {\langle \Psi | N(t) \Psi \rangle} \equiv \frac {A_0}{ \|\lambda\|^2}\,
  \lim_{t\to +\infty} \frac 1 t \, \langle n(\theta,\phi;t) \rangle\,,
\end{equation}
where $A_0$ is a kinematical factor to be determined and with dimensions of an area. To
determine $A_0$ let us consider the cross section for purely elastic scattering, for which the
Bohr--Peierls--Placzek formula (or optical theorem) gives $\sigma^{{\rm el}}(\theta,\phi) =
|q(\theta)|^2$, $\sigma^{{\rm el}}_{{}_{\rm TOT}}= 2 \, \frac {2\pi c } \omega\, \Im\, q(0)$;
the total cross section is the integral of the differential one on the whole solid angle. In
our case we have $\sigma^{{\rm el}}(\theta,\phi)=A_0
\left| \left( \left( S^\pm-\bbbone \right) \lambda \right) (\theta,\phi)\right|^2/
\|\lambda\|^2$ and, by the unitarity of $S^\pm$,
\[
\sigma^{{\rm el}}_{{}_{\rm TOT}}= \frac{A_0}{\|\lambda\|^2} \left\| \left( S^\pm -\bbbone
\right)\lambda\right\|^2 ={}-\frac{2A_0\sqrt\pi\, \Delta\theta}{\|\lambda\|} \, \Im\, \im
\e^{-\im\delta} \left( \left( S^\pm -\bbbone \right)\lambda \right)(0)\,.
\]
Then, by imposing the optical theorem, we get $A_0= \left(\frac{2\pi c}{\omega}\right)^2\frac1
{\pi(\Delta\theta)^2}$. Up to now we have not taken into account the polarization degrees of
freedom. If they are taken into account and the cross section for not polarized light is
considered, a $3/2$ extra--factor is obtained (\cite{CT92} pp.~532--533) and eq.~(\ref{3.19})
becomes
\begin{equation}\label{3.20}
  \sigma(\theta,\phi) = \left(\frac{2\pi c}{\omega}\right)^2\frac 3
{2\pi\eta^2} \,
  \lim_{t\to +\infty} \frac 1 t \, \langle n(\theta,\phi;t) \rangle\,.
\end{equation}

By differentiating eq.~(\ref{3.8}), we obtain
\begin{equation}\label{3.23}
  \frac{\dd \ }{\dd t} \left\langle n(\theta,\phi;t)\right\rangle = \Tr_{{}_\Hscr} \left\{
  R(\theta,\phi)^*\,R(\theta,\phi)\, \widetilde \rho_\lambda (t) \right\},
\end{equation}
\begin{equation}\label{3.24}
  R(\theta,\phi):= \alpha(\theta,\phi)\sigma_- + \big((S^+-\bbbone)\lambda\big)(\theta,\phi)
  P_+  + \big((S^--\bbbone)\lambda\big)(\theta,\phi) P_-\,;
\end{equation}
moreover, from eqs.~(\ref{3.23}) and (\ref{3.20}) we get
\begin{equation}\label{3.25}
  \sigma(\theta,\phi) = \left(\frac{2\pi c}{\omega}\right)^2\frac 3{2\pi\eta^2} \,
  \Tr_{{}_\Hscr} \left\{ R(\theta,\phi)^*\,R(\theta,\phi)\, \rho^\lambda_{\rm eq} \right\}.
\end{equation}
Finally, by eqs.~(\ref{2.21}), (\ref{3.12}), (\ref{3.13}), (\ref{3.25}), we obtain the
differential cross section and, by integrating it, the total one:
\begin{eqnarray}
  \sigma(\theta,\phi) &=& \frac 3 {2\pi \eta^2} \left( \frac{2\pi c} \omega \right)^2 \biggl\{
  \left( \frac{\|\alpha\|^2}{ 4\pi} + \eta^2 \left| g_+(\theta) + \frac 1 {4\sqrt \pi} \left(
  \e^{\im s_+}- 1 \right) \right|^2 \right) u(\infty)+{}\nonumber\\
\label{3.14} &+&\eta^2 \left| g_-(\theta) + \frac 1 {4\sqrt \pi} \left(
  \e^{\im s_-}- 1 \right) \right|^2 \bigl(1- u(\infty)\bigr)+{}\\
  &+&\frac{ \eta \|\alpha\|}{ \sqrt \pi} \, \Re \,\e^{-\im \delta} v(\infty) \left[ g_-(\theta)
  + \frac1 {4\sqrt \pi} \left( \e^{-\im s_-}- 1 \right) \right] \biggr\},\nonumber
\end{eqnarray}
\begin{equation}\label{3.15}
  \sigma_{{}_{\rm TOT}} = \frac 3 {2\pi } \left( \frac{2\pi c} \omega \right)^2 \biggl\{
  \left[ \| g_-\|^2 + \frac 1 2 \left( 1 - \cos s_-\right)  \right]\big(1- u(\infty)\big)+
u(\infty)\times {}
\end{equation}
\[
{}\times\left[ \| g_+\|^2 + \frac 1 2 \left( 1 - \cos s_+\right)  +
\frac{\|\alpha\|^2}{ \eta^2}\,\cos s_- -\frac{2\|\alpha\|^2 \Delta \omega}{ \eta^2 \kappa^2}\,
\sin s_- +
\frac{\|\alpha\|^2}{ 2 \kappa^2}\, \sin (\Delta s) \sin s_-\right]   \biggr\}.
\]

Let us end with some comments about $ \sigma_{{}_{\rm TOT}}$. When the elastic scattering is
negligible, i.e. when $g_\pm=0$ and $s_\pm=0$, eq.~(\ref{3.15}) reduces to
\begin{equation}\label{3.21}
   \sigma_{{}_{\rm TOT}} = \frac 3 {2\pi } \left( \frac{2\pi c} \omega \right)^2 \, \frac
   {\Gamma^2/4}{(\Delta \omega)^2 + \Gamma^2/4}\ \frac 1 {1+ 2 \eta^2/\|\alpha\|^2}\,,
\end{equation}
with $\Gamma^2 = \|\alpha\|^4 + 2 \|\alpha\|^2 \eta^2$ and $ \Delta \omega= \omega - \omega_0$.
For a laser with negligible intensity, i.e. when $\eta \downarrow 0$, eq.~(\ref{3.21}) reduces
to the cross section for resonant scattering, given in \cite{CT92} pp.~530--533; for $\eta \neq
0$, we have a power broadening of the resonance line, which maintains a Loretzian shape
\cite{Eze}.

To simplify the general case, we set
\begin{equation}\label{3.17}
  x:= \frac{2\Delta \omega}{\Gamma}\,, \qquad A:= \|g_-\|^2 + \frac 12 \left( 1- \cos s_-
  \right), \qquad C:={} - \frac{\|\alpha\|^2} \Gamma\, \sin s_-\,,
\end{equation}
\begin{eqnarray}
  B:=& &\!\!\!\!\!\!\! \frac{\eta^2 \kappa^2}{\Gamma^2} \left( \|g_+\|^2 - \|g_-\|^2 +
  \frac 12 \,\cos s_- -  \frac 12\, \cos s_+\right) +{}\nonumber\\
{}+ & & \!\!\!\!\!\!\!  \frac {\|\alpha\|^2 }{\Gamma^2} \left[ \kappa^2 \cos s_- +
  \frac{\eta^2}2 \,  \sin(\Delta s)\sin s_- \right].
\label{3.18}\end{eqnarray}
Then, we can write
\begin{equation}\label{3.16}
  \frac{2\pi} 3 \left(\frac \omega {2\pi c}\right)^2 \sigma_{{}_{\rm TOT}} = A+
  \frac{B+ C x}{x^2 +1}\,;
\end{equation}
the positivity of $\sigma_{{}_{\rm TOT}}$ is equivalent to $A> 0 $ and $ A(A+B)\geq C^2/4 $ or
$A=0$, $B>0$, $C=0$. According to the values of the various coefficients and mainly to the
signs of $B$ and $C$, different line shapes appear, which are known as Fano profiles
(\cite{CT92} pp.~61--63). These shapes are typical of the interference among various channels,
when one of them has an amplitude with a pole in the complex energy plane; in our case the
channels are elastic scattering in the up state, elastic scattering in the down state and
fluorescence.

Whichever be the line shape, there is a strong variation of the cross section for $\omega$
around $\omega_0 +\epsilon$,
\begin{equation}\label{3.26}
  \epsilon:= \eta^2\, \Im\, \langle g_+| g_- \rangle + \sqrt
  \pi\, \eta^2 \,\Im\, \Delta g(0)\,.
\end{equation}
The intensity dependent shift $\epsilon$ of the resonance frequency has received various names
in the literature; a very suggestive one is \emph{lamp shift}, a name suggested by A.~Kastler
in \cite{Kast}. Note that in our two--level system the lamp shift is not vanishing only if the
two states respond differently to elastic scattering; moreover, only the not $s$-wave
contribution does matter. Let us stress that also the line width $\Gamma$ and the whole line
shape are intensity dependent.

\end{document}